\documentclass[twocolumn,prb,showpacs,multicol,amsmath,amssymb]{revtex4}
\usepackage[dvips]{graphicx}
\usepackage{amsmath}
\usepackage{graphicx}
\usepackage{dcolumn}
\usepackage{bm}
\usepackage{graphics}
\usepackage{soul}
\usepackage{epsfig,color}
\usepackage[left]{lineno}
\usepackage{blindtext}
 \usepackage{relsize}

\newcommand{\be}{\begin{equation}}
\newcommand{\ee}{\end{equation}}
\newcommand{\bea}{\begin{eqnarray}}
\newcommand{\eea}{\end{eqnarray}}

\definecolor{airforceblue}{rgb}{0.36, 0.54, 0.66}
\definecolor{brightgreen}{rgb}{0.4, 1.0, 0.0}
\definecolor{cadmiumgreen}{rgb}{0.0, 0.42, 0.24}
\definecolor{canaryyellow}{rgb}{1.0, 0.94, 0.0}
\definecolor{capri}{rgb}{0.0, 0.75, 1.0}
\begin{document}
\title[J. Vahedi ]{Spin and charge thermopower effects in the ferromagnetic graphene junction}
\author{Javad Vahedi$^{1,2}\footnote{email: javahedi@gmail.com\\ Tel: (+98) 911-1554504\\Fax: (+98) 11-33251506}$ and  Fattaneh Barimani$^{2}$ }
\address{$^{1}$Department of Physics, Sari Branch, Islamic Azad University, Sari, Iran.}
\address{$^{2}$Center for Theoretical Physics of Complex Systems, Institute for Basic Science (IBS), Daejeon, Korea.}
\date{\today}
\begin{abstract} 
Using wave function matching approach and employing the Landauer-Buttiker formula a ferromagnetic graphene junction with temperature gradient across the system, is studied. We calculate the thermally induced charge and spin current as well as the thermoelectric voltage (Seebeck effect) in the linear and nonlinear regimes. Our calculation revealed that owing to the electron-hole symmetry, the charge Seebeck coefficient is, for an undoped magnetic graphene,  an odd function of chemical potential while the spin Seebeck coefficient is an even function regardless of the temperature gradient and junction length.  We have also found with an accurate tuning external parameter, namely the exchange filed and gate voltage, the temperature gradient across the junction drives a pure spin current without accompanying the charge current. Another important characteristic of thermoelectric transport, thermally induced current in the nonlinear regime, is examined.  It would be our main finding that with increasing thermal gradient applied to the junction the spin and charge thermovoltages decrease and even become zero for non zero temperature bias. 
 \end{abstract}

\maketitle


\section{INTRODUCTION}\label{sec1}
There is a fast growing attention to graphene because it has a rich potential not only from the fundamental side but also from the applied point of view\cite{a1,a2,a3}. Graphene is a single layer of carbon atoms arranged in a two-dimensional honeycomb lattice\cite{a4,a5}. The study of its electronic properties has recently found great interest\cite{a6,a7,a8,a9} in part owing to the peculiar features of its energy bandstructure. Within a tight-binding model, graphene's  valence and conduction bands touch each other at six different points, the K-points, which reduce to two, K and $K^\prime$, because the symmetry analysis show the rest are equivalent. Near these points and at low excitations, electrons behave as massless fermions traveling at fixed velocity $v_F\sim10^6m/s$, independent  of their energy.  Graphene has many important features of applications: it shows gate-voltage-controlled carrier conduction, high field-effect mobilities and a small spin-orbit interaction.\cite{a10,a11}.
 \begin{figure}
 \includegraphics[width=1.05\columnwidth]{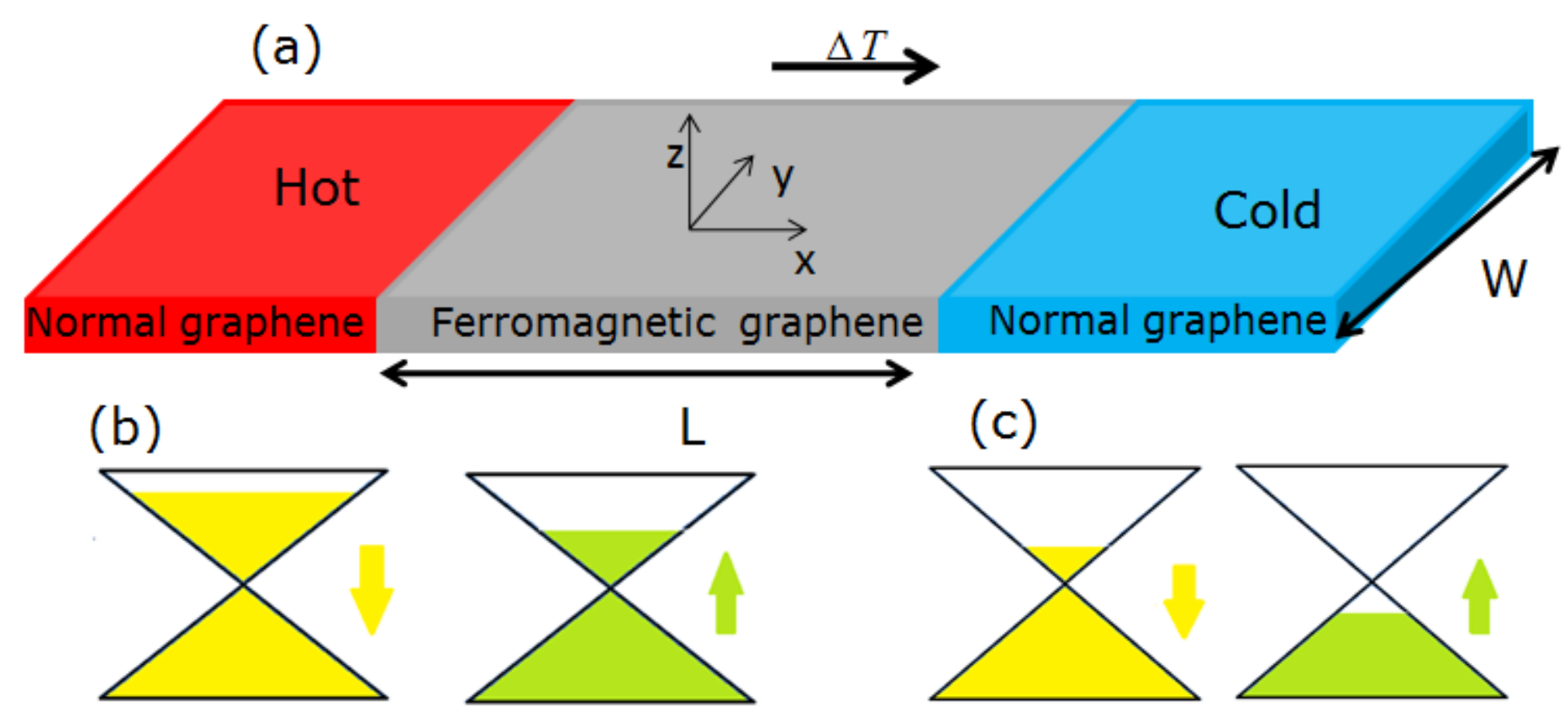}
 \caption{(Color online) (a) A schematic diagram of normal$/$ferromagnetic$/$normal graphene junction. The length of the central  region, the ferromagnetic graphene, is $L$. Electronic transport is activated with a temperature gradient $\Delta T$ between the two hot and cold electrodes. (b) and (c) are pictorial illustration of two up and down spin subbands shift of the magnetic graphene. (b) The two spin subbands are located in the conduction band. (c) One of the spin subband, here up spin, is located in the conduction band and the down spin is shifted to the valance band.}
\label{fig1}
\end{figure} 
\par
In recent years, the well known thermoelectric effects have gain great of attention due to their crucial relevance in mesoscopic and nanoscopic systems.\cite{a12,a13}. On the one hand, the studies can be helpful technologically in managing the generated heat in nanoelectronic devices.  On the other hand, investigations about thermoelectric effects in mesoscopic regimes are of fundamental interest for condensed matter physicists. The thermoelectric properties of graphene have been also studied both experimentally and theoretically with special focus on the charge neutrality (the Dirac) point\cite{a14,a15,a16,a17,a18,a19,a20,a200,a21,a211,a2111,a22}. One of the key findings has been the sign change of the thermoelectric power across the Dirac point when the carriers type switches from electron to hole, accompanied by the enhancement behavior of Seebeck coefficient.\cite{a17}.
\par
Beginning in the late 1980’s the field of spintronics emerged which focuses on the characteristic of spin-dependent transport  and its coupling to the charge\cite{a23,a24,a25}. Along with the fast developing and high demanding  interest in this field, the seminal work of Johnson and Silsbee displayed that, in spintronic and magnetic devices, heat currents can couple to the spin currents and also the charge currents\cite{a26}. Moreover,  some succeeding unexpected experimental achievements of spin Seebeck effects\cite{a27,a28,a29,a30}, attracted great attention to study the thermoelectric and spintronic effects together which steer to the introduction of a new research field, spin caloritronics\cite{a31,a32}.
\par
In this paper, we consider normal/ferromagnetic/normal graphene junctions where a gate electrode is attached to the ferromagnetic graphene (see Fig.~\ref{fig1}-a). 
We study the combination of charge, heat and spin transport in graphene in the context of spin caloritronics and spin-dependent thermoelectric phenomena. We have  found with an accurate tuning external parameter, namely the exchange filed ($H$) and gate voltage (U), the temperature gradient across the junction drives a pure spin current without accompanying the charge current. 
\par
In Fig.~\ref{fig1}-(b) and (c), we show schematically two possible situations, namely $U>H$ and $U<H$, respectively. Depends on the gate voltage magnitude (throughout the present work, we choose $U\geq 0$), the Fermi energy level (both degenerate  up and down spin subbands) shifts away from the neutrality point to the conduction band. So the competition between exchange filed and gate voltage plays an important role, in which tuning the exchange filed shifts the two up and down spin subbands in different ways and two possible situations are predicted.
\par
The paper is organized as follows. In section (II), we summarize the model, Hamiltonian and formalism. In section (III), we present our numerical results. Finally, conclusion is given in section IV.
\\
\section{Computational Scheme} \label{sec2}
The fermions around Fermi level in graphene can be defined by  a massless relativistic Dirac equation. The Hamiltonian is given by
\begin{equation}
\emph{H}_{\pm}=v_{F}(\rho_{x}k_{x}\pm\rho_{y}k_{y})
\label{e1}
\end{equation}
 with the Pauli matrices $\rho_{x}$ and $\rho_{y}$ and the velocity $v_{F}\sim{10}^{6}m/s$ in graphene. The Pauli matrices act on the two sublattice of the honeycomb structure. The $\pm$ sign refers to the two valleys of K and $K^\prime$ points in the first Brillouin zone. Moreover,  the valley degeneracy permits one to consider one of the ${H}_{\pm}$ set. A two dimensional normal/ferromagnetic/normal graphene junction is considered where an external transverse electric field is applied to a part of graphene sheet to make it ferromagnetic partially\cite{a33}.  Using the first-principles calculations authors in ref.[\cite{a34}] has shown that an applied in-plane homogeneous electric fields across the graphene nanoribbons, can induce half-metallic properties.  Besides the applied electric field, recently it has been shown that placing graphene on an insulating ferromagnetic substrate made of yttrium iron garnet (YIG) can make the graphene ferromagnetic while leaving its electronic properties unchanged\cite{a34-1}. A gate electrode is also attached to the ferromagnetic graphene.
\par
 The interfaces are parallel to the y-axis and located at $x=0$ and $x=L$  (see Fig.~\ref{fig1}-a). Since there is a valley degeneracy, one can focus on the Hamiltionian $H_{+}$ with $H_{+}=v_{F}(\rho_{x}k_{x}+\rho_{y}k_{y})-V(x)$, where $V(x)=\mu_{F}$ in the normal graphenes and $V(x)=\mu_{F}+U-\sigma H$ in the ferromagnetic graphene. Here, $\mu_{F}=v_{F}k_{F}$ is the Fermi energy, $U$ is the chemical potential shift tunable by the gate voltage, and $H$ is the exchange field. $\sigma=\pm$ signs correspond to majority and minority spins. The spin dependent band then follows as $\varepsilon_\sigma=v_F(k-k_F)-U-\sigma H$. The wave-functions are given by
\begin{equation}
\Psi_L=\Psi_L^{+}+a_{\pm}\Psi_L^{-},\quad\Psi_M=b_{\pm}\Psi_L^{+}+c_{\pm}\Psi_M^{-},
\Psi_R=d_{\pm}\Psi_R^{+}
\label{e2} 
\end{equation}
with
\begin{eqnarray}
\Psi_L^{\pm}&=&\begin{pmatrix}1\\ \pm e^{\pm i\theta}\end{pmatrix} e^{\pm ipx\cos\theta+ip_yy}\nonumber\\
\Psi_M^{\pm}&=&\begin{pmatrix}1\\ \pm e^{\pm i\theta^\prime}\end{pmatrix} e^{\pm ip\prime x\cos\theta^\prime+ip_yy}\nonumber\\
\Psi_R^{\pm}&=&\begin{pmatrix}1\\ e^{\pm i\theta}\end{pmatrix} e^{\pm ipx\cos\theta+ip_yy}
\end{eqnarray}
where $\psi_{L,(R)}$ demonstrates the wave-function in the left (right) normal graphene while $\Psi_M$ is the wave-function in the ferromagnetic graphene, with angles of incidence $\theta$ and $\theta^\prime, p=(E+\mu_{F})/v_{F}$ and $ p_{\pm}^\prime = (E+\mu_{F}+U \pm{H})/v_{F}$. Using the translational symmetry in the y-direction of the junction, one can show the momentum parallel to the y-axis is conserved: $p_y=p\sin\theta = p\prime\sin\theta^\prime$.
\par
By matching the wave functions at the interfaces$ \Psi_L=\Psi_M$ at  $x=0$ and $\Psi_M =\Psi_R$   at  $x=L$,  we obtain the coefficients in the wave-functions. The transmission coefficient has the form 
\begin{equation}
d_\sigma=\frac{e^{- ipl\cos\theta}}{\cos(p_\sigma^{\prime}l\cos\theta^{\prime})-i\sin(p_\sigma^{\prime}l\cos\theta^{\prime})[\frac{1-\sin\theta\sin\theta^\prime}{\cos\theta\cos\theta^\prime }]}\nonumber
\label{e3}
\end{equation}
which can be used to calculate the total transmission probability $\mathlarger{\mathlarger{\mathlarger{\tau}}}_\sigma=|d_\sigma|^2$.
Now having the transmission probability, the current can be written as 
\begin{equation}
I_\sigma=\frac{e}{\hbar}\int dEN(E) \int d\theta \cos\theta \mathlarger{\mathlarger{\mathlarger{\tau}}}_\sigma \Big[f_{L\sigma}(E)-f_{R\sigma}(E) \Big]
\end{equation}
where $ N(E)=\frac{|E-\mu_F|W}{\hbar v_F}$,  W is the width of the graphene sheet, is the carrier density of state (DOS).  $f_{\alpha\sigma}(E)=\frac{1}{1+e^{[(E-\mu_{\alpha\sigma})/k_B T_\alpha ]}}$ with $\alpha=L,R$ are the Fermi-Dirac distribution function in each contact with spin $\sigma$. Defining the electrochemical potential as $\mu_{\alpha\sigma}=\mu_F+eV_{\alpha\sigma}$ where  $V_{\alpha\sigma}$ is the voltage in contact $\alpha$ with spin $\sigma$ which accounts for possible population imbalances between different spin subbands. 
\begin{figure}
 \includegraphics[width=1.05\columnwidth]{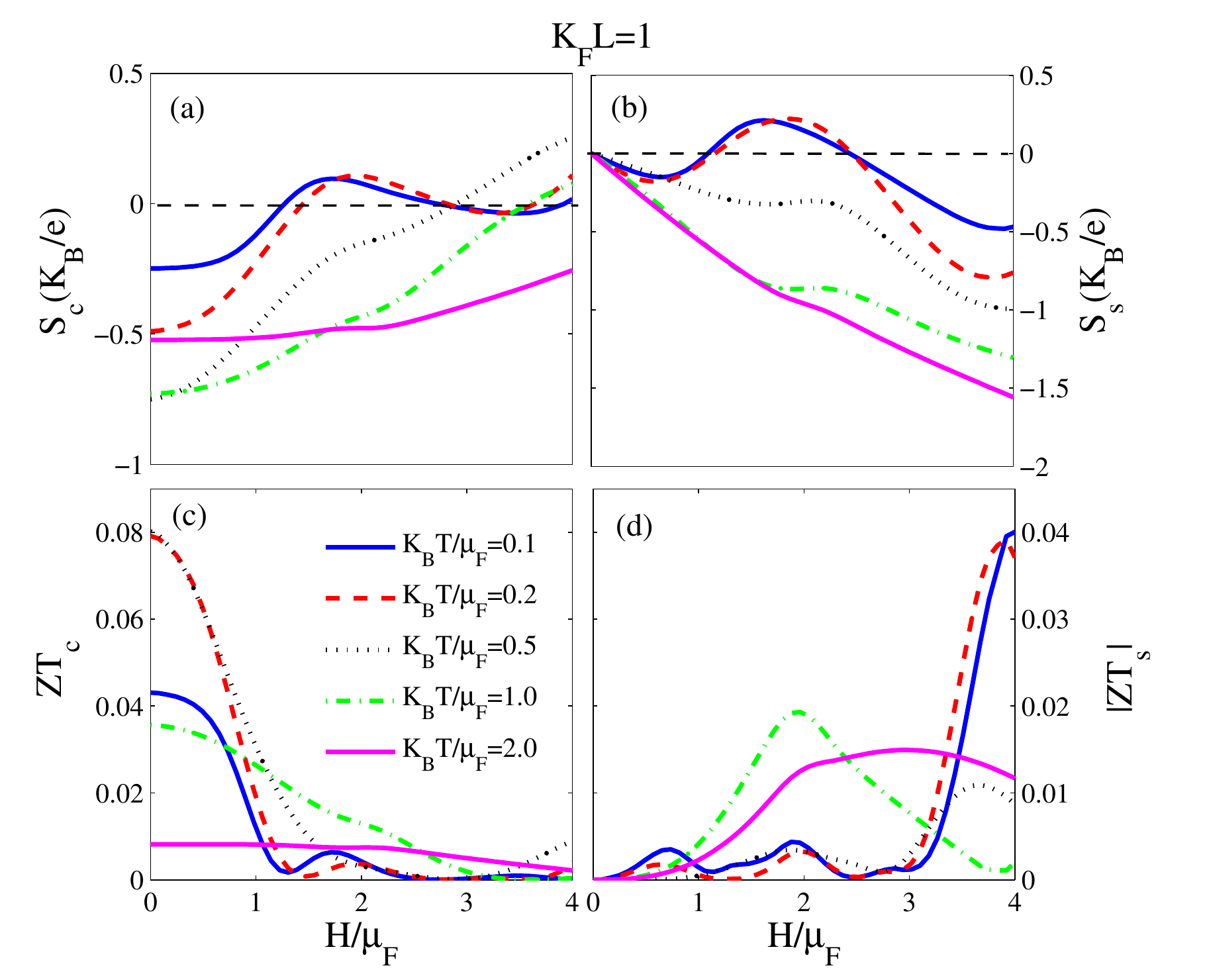}
 \caption{(Color online) Charge and spin thermopower (left and right top panels, respectively) and corresponding figures of merit (left and right bottom panels, respectively)  are given as a function of exchange field $ H/\mu_F$ for different $k_BT/\mu_F$. We fix other parameters as $k_FL=1$ and $U/\mu_F=2$. The Seebeck coefficient is measured in units of ($k_B/e$).}
\label{fig2}
\end{figure}
\begin{figure}
\includegraphics[width=1.05\columnwidth]{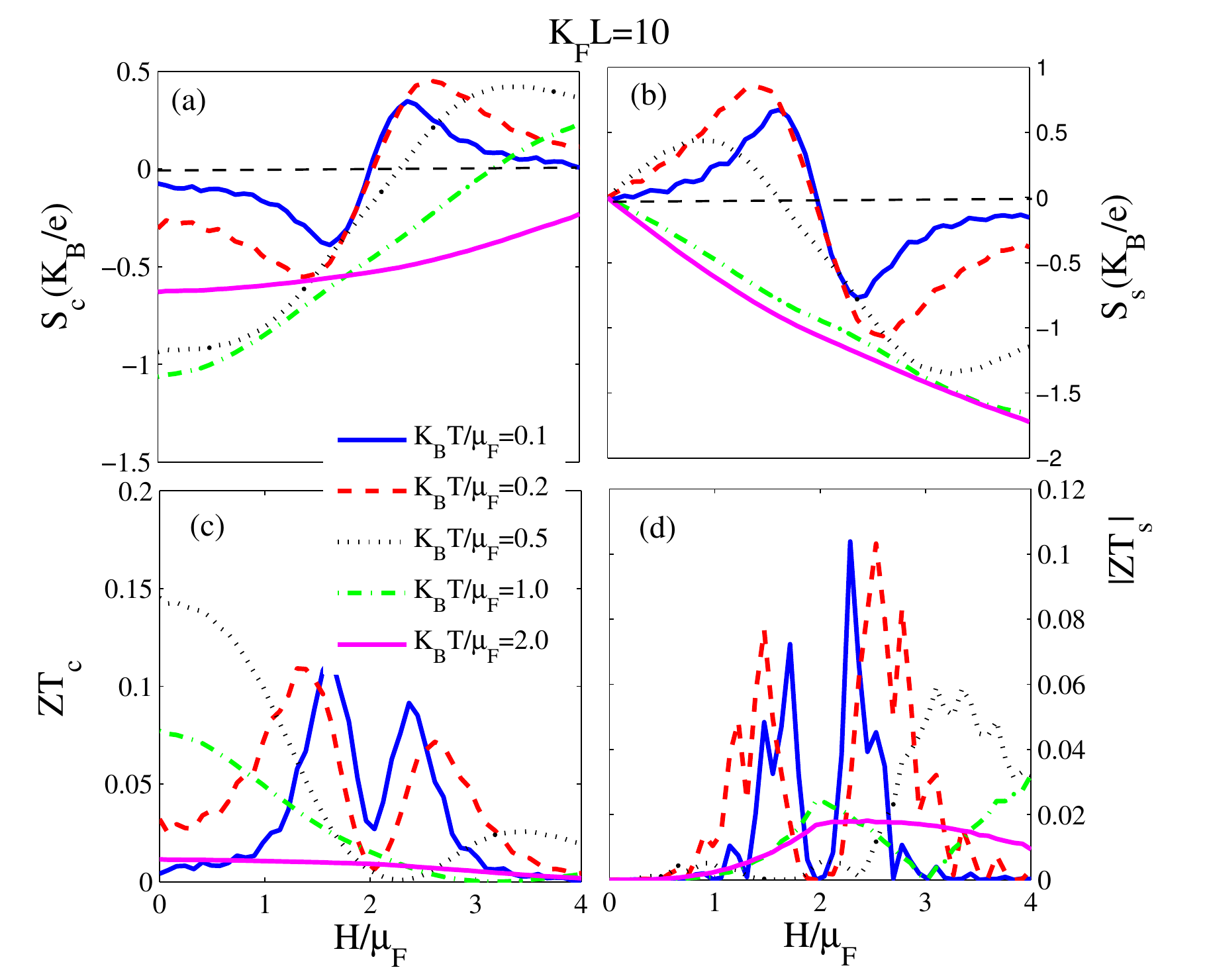}
 \caption{(Color online) Charge and spin thermopower (left and right top panels, respectively) and corresponding figures of merit (left and right bottom panels, respectively)  are given as a function of exchange field $ H/\mu_F$ for different $k_BT/\mu_F$. We fix other parameters as $k_FL=10$ and $U/\mu_F=2$. The Seebeck coefficients are measured in units of  ($k_B/e$).}
\label{fig3}
\end{figure}
\par
Employing the linear response assumption, i.e.,$ T_{L}\approx T_{R}=T$, we obtain the spin resolved thermopower $$ S_{\sigma}=-\frac{1}{eT}\frac{L_{1,\sigma}}{ L_{0, \sigma}}$$ with $L_{n(=0,1)\sigma}=\frac{1}{\hbar}\int dE (E-\mu)^n N(E) \int d\alpha\cos\alpha T_{\sigma}\big[-\partial_{E}f(E)\big]$. The charge and spin thermopowers $S_{c}$ and $S_{s}$ are calculated as $S_{c}=\frac{S_{+}+S_{-}}{2}$ and $S_{s}=S_{+}-S_{-}$. The charge and spin figures of merit for a magnetic system can be defined versus Seebeck coefficients  as below, $$Z_{ch(sp)}T=\frac{G_{ch(sp)}S_{ch(sp)}^2T}{\kappa}$$ where $G_{ch}=G_{+}+G_{-}(G_{sp}=|G_{+}-G_{-}|)$ denotes charge(spin) conductivity with $G_\sigma=e^2L_{0,\sigma}$  and the thermal conductivity is given by $\kappa=\kappa_{+}+\kappa_{-}$. We concentrate in low enough temperatures where only electrons contribute effectively in thermal transport and at this regime spin dependent $k_{\sigma}$ reads by definition, $$\kappa_{\sigma}=\frac{1}{T}\Big(L_{2,\sigma}-\frac{L_{1,\sigma}^2}{L_{0,\sigma}}\Big)$$
In this work, we will not consider the phonon contribution in the thermal conductivity. The  main contribution of $\kappa_{ph}$ would lead to a smaller figures of merit because it enhances the denominator of $ZT$. So the charge and spin thermopower will not be affected with the presence of phonon. Furthermore, It has been reported that at low temperature the thermal conductivity of phonon decreases with temperature in a power-law fashion $(\kappa_{ph}\propto T^{1.68})$, while the thermal conductivity of electrons shows a linear behavior $(\kappa_{el}\propto T)$ \cite{a34n,a35n}. Moreover, in Ref.[\cite{Lindsay14}],  has been shown that  the thermal conductivity of phonon decreases with length.  It is also worth mentioning  that the temperature dependent have been reported for graphene with $L=10\mu m$ which is at least ten times bigger than what we consider in this work. So at low temperature ($T\le10K$) and with length ($\le 1\mu m$) the thermal conductivity of electrons has a dominant contribution than the thermal conductivity of phonon. 
\section{Numerical Results} \label{sec3}
We present the numerical results for both linear and nonlinear regimes separately.  When $\Delta\theta=T_L-T_R\ll T_{L,R}$ ($T_{L,R}$ are the temperatures of the left and right electrode, respectively) the system acts within the linear in temperature regime. In this regime, the thermopower characterizes the efficiency of energy conversion along with the thermoelectric figure of merit $ZT$. While the temperature differential between electrodes enhances, the system may shift to nonlinear regime of action. For example, a thermovoltage that nonlinearly alter with $\Delta\theta$ was reported experimentally and theoretically on semiconductor quantum dots and single-molecule junctions\cite{a36n,a37n,a38n}. We focus on the charge and spin Seebeck coefficients $(S_{ch}, S_{sp})$ and their corresponding figures of merit $ Z_{ch}T$  and$ Z_{sp}T$. It is widely known that electron-hole asymmetry near the Fermi level in the band structure or transport properties controls  the thermoelectric effects. So one envisages that manipulating such asymmetry would be applicable to find such a eminent thermoelectric effect in graphene junction. In the following, we first address the linear regime. 
\subsection{Linear regime}
In the calculations, all energy scaled with the Fermi energy $\mu_F$ and we set it as the unit of energy. Fig.~\ref{fig2} shows the charge and spin thermopower (panels (a) and (b), respectively) and corresponding figures of merit (panels (c) and (d), respectively) as functions of scaled exchange field $H/\mu_F$ for different scaled temperature $k_BT/\mu_F$. We have set dimensionless gate voltage $U/\mu_F=2$, in which the Fermi level lies in the conduction band. As it can bee seen, at low temperature limit, here we mean $k_BT/\mu_F=0.1$ and $0.2$, both charge and spin thermopowers present an oscillation trend as a function of exchange field. At zero exchange field, both $S_{+}$ and $S_{-}$ have an equal and negative contribution, so one expects a zero spin Seebeck coefficient $S_{sp}=0$, which is clear from the result depicted in  Fig.~\ref{fig2}-(b). The total negative charge Seebeck coefficient $S_{ch}<0$ is owing to a charge accumulation gradient in the opposite direction of moving electrons in the conduction band with both up and down spins along the temperature gradient.
\begin{figure}
  \includegraphics[width=1.07\columnwidth]{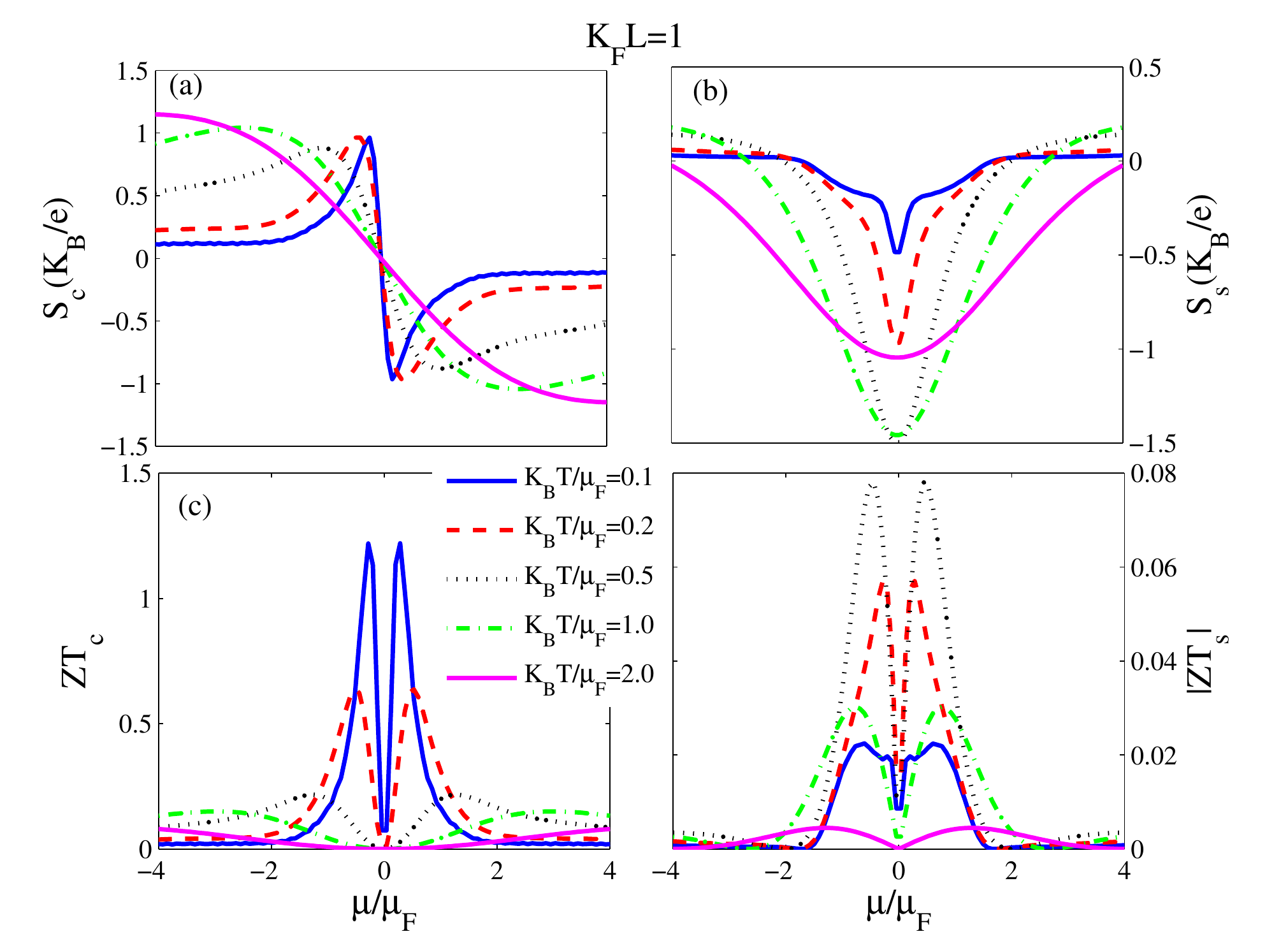}
 \caption{(Color online) Charge and spin thermopower (left and right top panels, respectively) and corresponding figures of merit (left and right bottom panels, respectively) are given as a function of $\mu/\mu_F$ for different values of dimensionless temperature $k_BT/\mu_F$. Here we set $U/\mu_F=0$, $H/\mu_F=2$, $k_FL=1$ and the Seebeck coefficients are measured in units of ($k_B/e$).}
\label{fig4}
\end{figure}
\begin{figure}
\includegraphics[width=1.07\columnwidth]{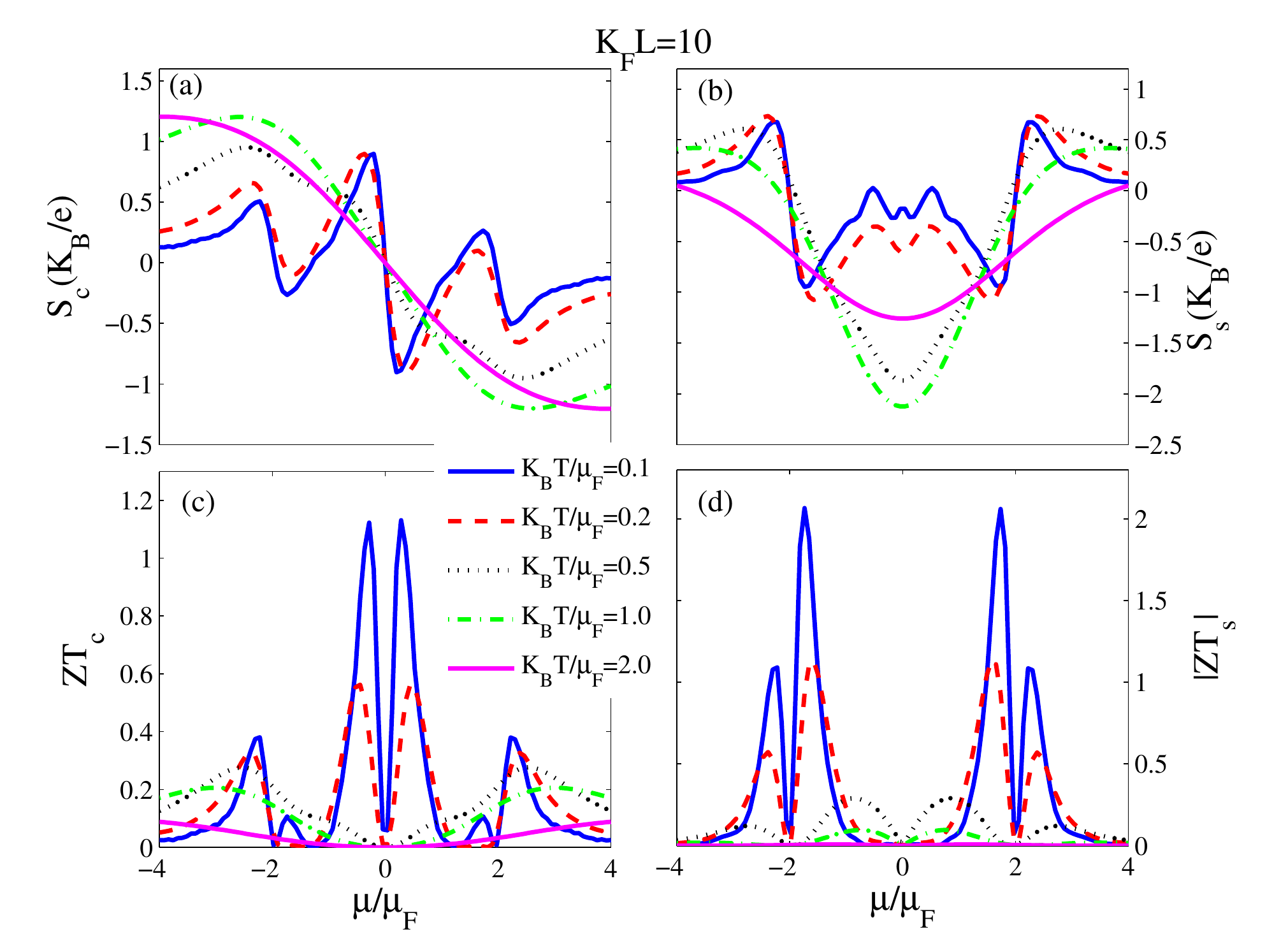}
 \caption{(Color online) Charge and spin thermopower  (left and right top panels, respectively) and corresponding figures of merit (left and right bottom panels, respectively) are given as a function of $\mu/\mu_F$ for different values of dimensionless temperature $k_BT/\mu_F$. Here we set $U/\mu_F=0$, $H/\mu_F=2$, $k_FL=10$ and the Seebeck coefficients are measured in units of ($k_B/e$).}
\label{fig5}
\end{figure}
\par
For all temperatures and exchange filed considered here, except in the region $1.5<H<2.5$ at low temperature, both charge and spin Seebeck coefficients are negative. It signals that even in the presence of the exchange filed the majority spin carries, here down spin, from the conductance band dominates. For the middle region $1.5\lesssim H/\mu_F \lesssim2.5$ at low temperature, the situation is reversed and the contribution of minority spin carriers from the valence band dominates.  In this region the exchange field shifts the up spin subband's Fermi level to the valence band and cause the holes from spin down subband to be thermally activated. These excitations transfer positive charge current and so, has positive sign contribution to the charge Seebeck while the moved spin up electrons to the conduction band yet have a negative contribution. It is also worth noting that at high temperatures, the spin Seebeck coefficient is negative for all exchange fields, while its charge counterpart still shows a changing sign. Which it can be regarded as a pure spin current caused by temperature gradient. 
\begin{figure*}
 \includegraphics[width=0.8\columnwidth]{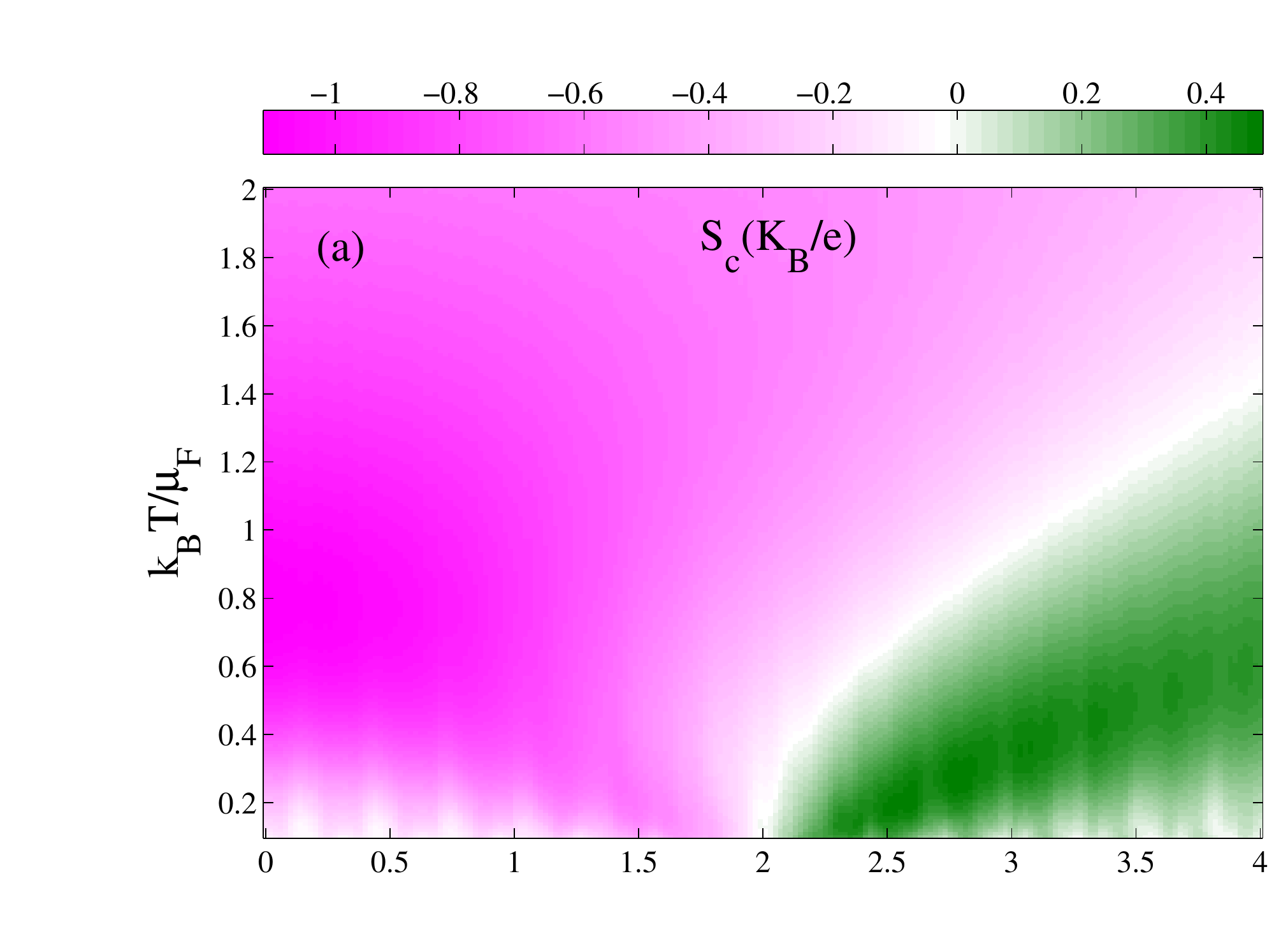}
\includegraphics[width=0.8\columnwidth]{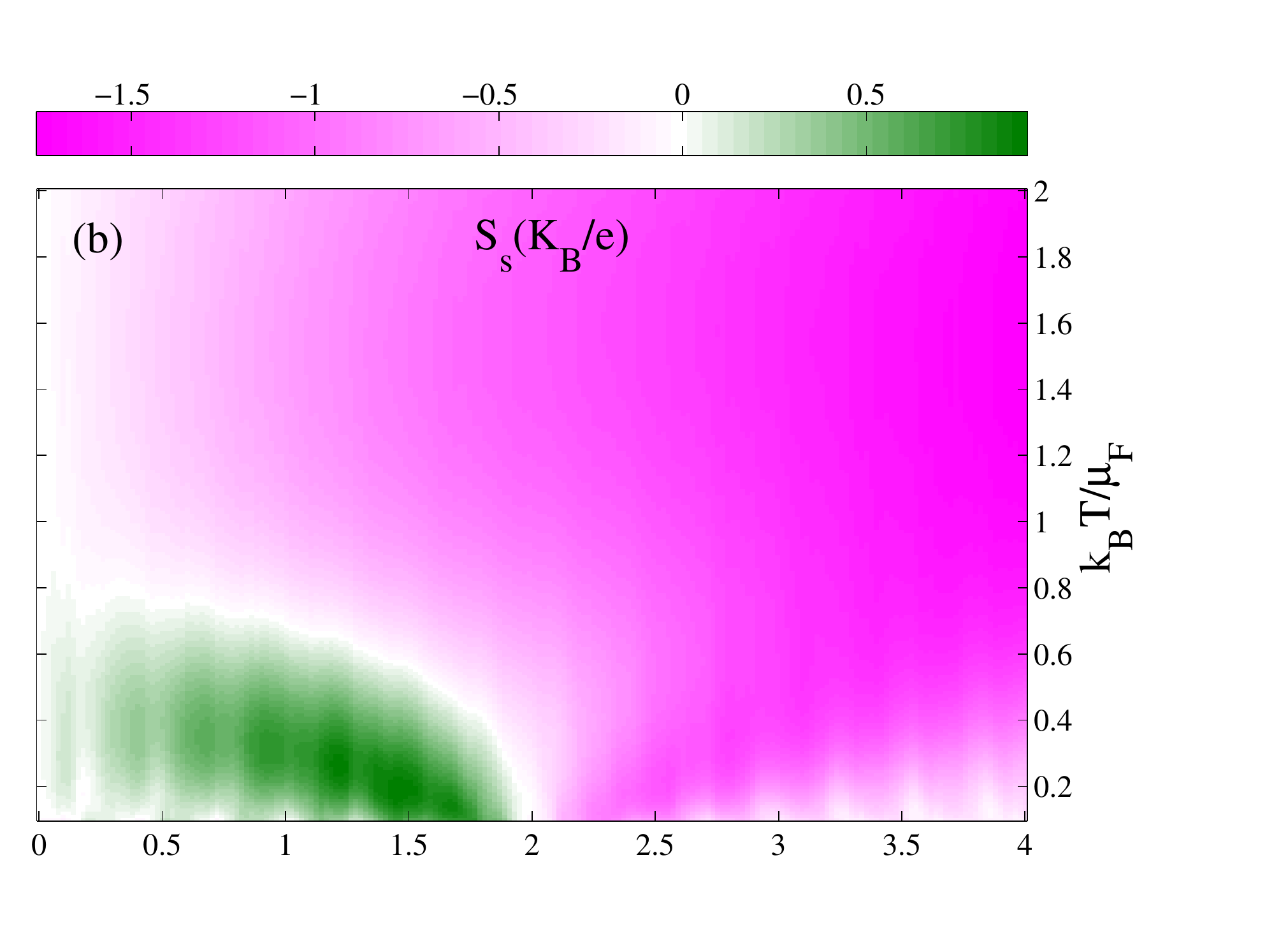}\\
\includegraphics[width=0.8\columnwidth]{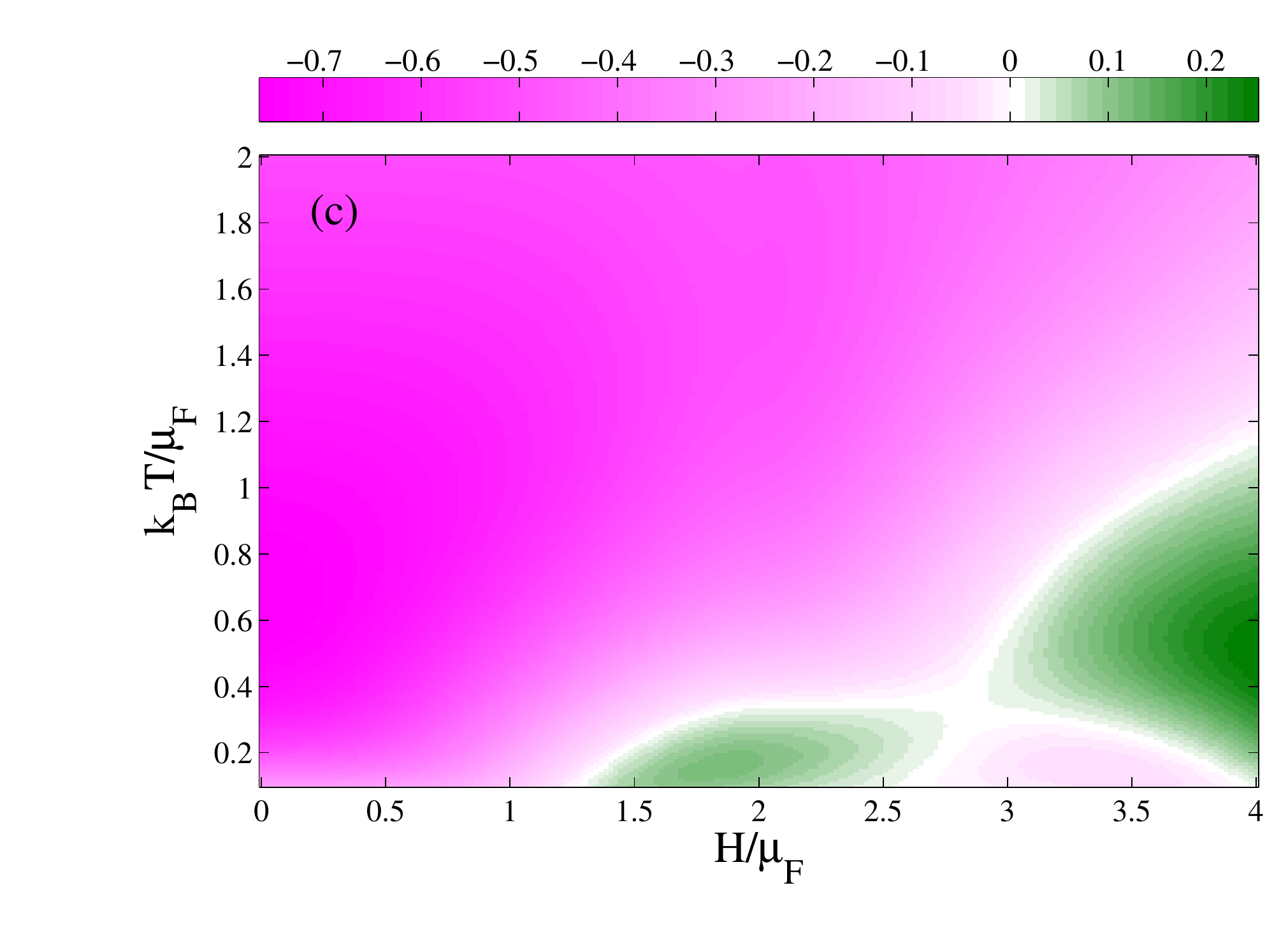}
\includegraphics[width=0.8\columnwidth]{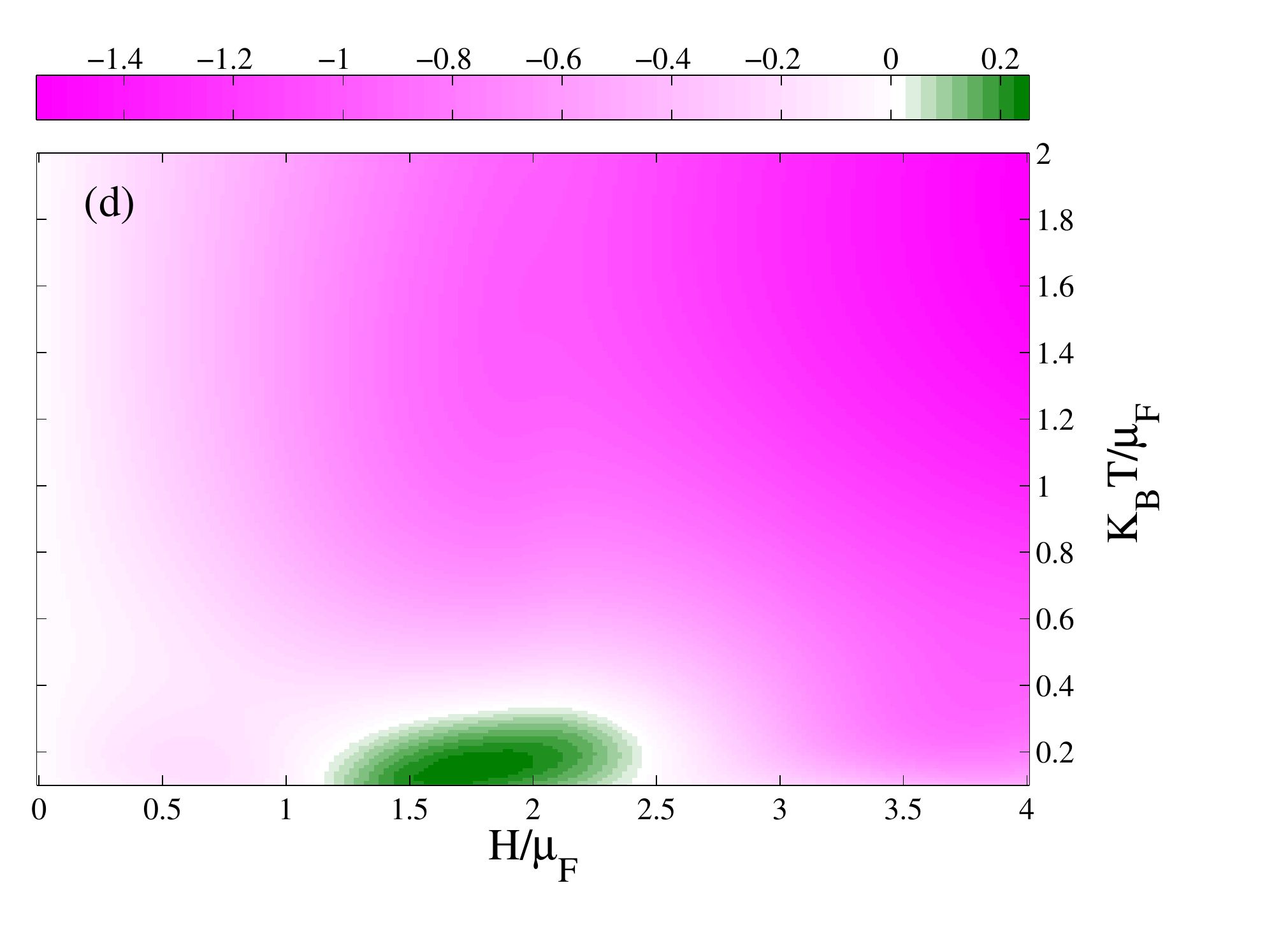}
 \caption{(Color online) Density plot of the charge (right column) and spin (left column) thermopowers versus dimensionless exchange field $H/\mu_F$ and temperature $k_BT/\mu_F$. Top and bottom panels are correspond to the $k_FL=10$ and $k_FL=1$, respectively.  The dimensionless gate voltage is set as: $U/\mu_F=2$.}
\label{fig6}
\end{figure*}
\par
To get more insight, in Fig.~\ref{fig3} we consider the effect of a different length. So as the previous case, here we have repeated our calculation with $k_FL=10$. The other parameters are same as the Fig.~\ref{fig2}. Compare with the previous case, $k_FL=1$, the profound effects are visible at low temperatures. Interestingly, both the charge and spin thermopower show an odd feature respect to the exchange field $S_{c(s)}(-H)=-S_{c(s)}(H)$ which reaches the maxima on one side of the symmetric point $H/\mu_F=U/\mu_F$ and the minima on the other side. The corresponding figure of merits also shows an enhancement about the symmetric point $H/\mu_F=2$  which is profound at the low temperature limit. In current experimental situation, it is feasible to find such a tunability of length in order to reach such enhancement in the power output of the thermoelectrical devices.
\par
Fig.~\ref{fig4} shows the charge and spin thermopowers and corresponding figures of merit as functions of scaled chemical potential $\mu/\mu_F$ for different values of dimensionless temperature $k_BT/\mu_F$ with length $k_FL=1$. The charge Seebeck coefficient shows an odd function of  $\mu/\mu_F$ in which reaches the maxima on one side of the symmetric point $\mu/\mu_F=0$ and the minima on the other side. While the spin Seebeck coefficient is an even function of  $\mu/\mu_F$ and can reach a minimum at $\mu/\mu_F=0$, where the charge Seebeck coefficient is zero, so it can be feasible to obtain a pure spin thermopower.  The physics behind the generation of a pure spin thermopower backs to symmetrical shifts of different spin up and down subband about $\mu/\mu_F=0$, (see Fig.~\ref{fig1}-c), in which $S_{+}(0)=-S_{-}(0)$. It is worth to mention that from the symmetry of spin-dependent band energy $\varepsilon_\sigma(\mu)=-\varepsilon_\sigma(-\mu)$, we have $S_+(\mu)=-S_-(-\mu)$ which leads to $S_c(\mu)=-S_c(-\mu)$ and $S_s(\mu)=S_s(-\mu)$. The large spin Seebeck coefficient is observed in the intermediate temperature, which even can exceed its charge counterpart in magnitude. 
\par
Effects of different length are also addressed in Fig.~\ref{fig5}.  It is clear, by comparing cases with lengths  $k_FL=1$ and $k_FL=10$, much more effect occurs at low temperature limit in which the pure spin thermopower finds a big reduction at the symmetric point $\mu/\mu_F=0$, meanwhile gets a big enhancement with sign changing about $\mu/\mu_F=\pm H/\mu_F$. A careful inspection reveals that at these two points $\mu/\mu_F=\pm2$, both $S_c$ and $S_s$ are zero, which can be achieved when $S_{+}(\pm 2)=S_{-}(\pm 2) =0$. It is also worth noticing that at these points the spin figure of merit $|ZT_s|$ finds magnitude more than $2$.
\par
Before closing the linear regime behavior and in order to have a comprehensive discussion, in Fig.~\ref{fig6} we have depicted a density plot of the charge and spin thermopower versus dimensionless exchange field $H/\mu_F$ and temperature $k_BT/\mu_F$. As is can be seen, the major effects happen in the low temperature regime. In case with $k_FL=10$ (see two top panels), at very low temperature $S_c$ ($S_s$) shows a sharp sign change from negative (positive) to positive (negative) at $H/\mu_F=U/\mu_F$. Increasing temperature shifts this point to higher and lower magnitude of the exchange field for the charge and spin thermopowers, respectively. The white color line in the figures is signaling zero thermopower which one can see with fine-tunability of background temperature, and exchange filed it would be feasible to find a large pure spin thermopower without accompanying  charge ones. For $k_FL=1$ case (see two bottom panels), the situation is less regular and at very low temperature  one can see two times which $S_c$ and $S_s$ change their sign.  It is interesting that contrast to the case with $k_FL=10$, $S_c$ and $S_s$ show the same sign at very low temperature. While by increasing temperature the situation gets more complex for the charge thermopower with zero magnitude which spin thermopower finds big negative value.
\subsection{Non-linear regime}
\begin{figure}
 \includegraphics[width=1.05\columnwidth]{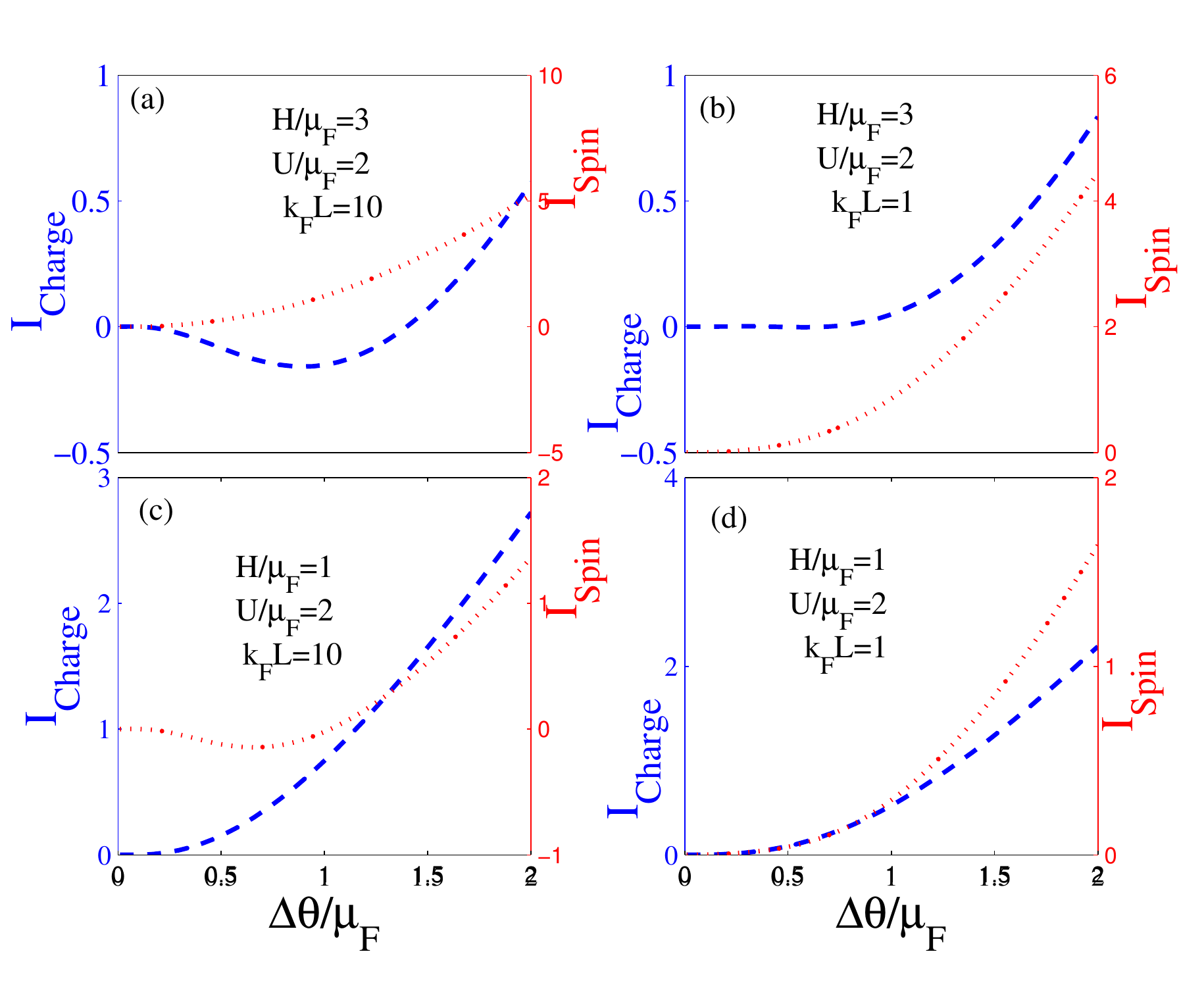}
 \caption{(Color online) Charge (blue dashed line) and spin (red dotted line) thermocurrent given as functions of temperature difference $\Delta\theta$ for a set of parameters which indicated in each panel. In all cases we fixed the dimensionless gate voltage as $U/\mu_F=2$. Right and left columns correspond to the $k_FL=10$ and $k_FL=1$ cases. }
\label{fig7}
\end{figure}

\begin{figure}
 \includegraphics[width=1.0\columnwidth]{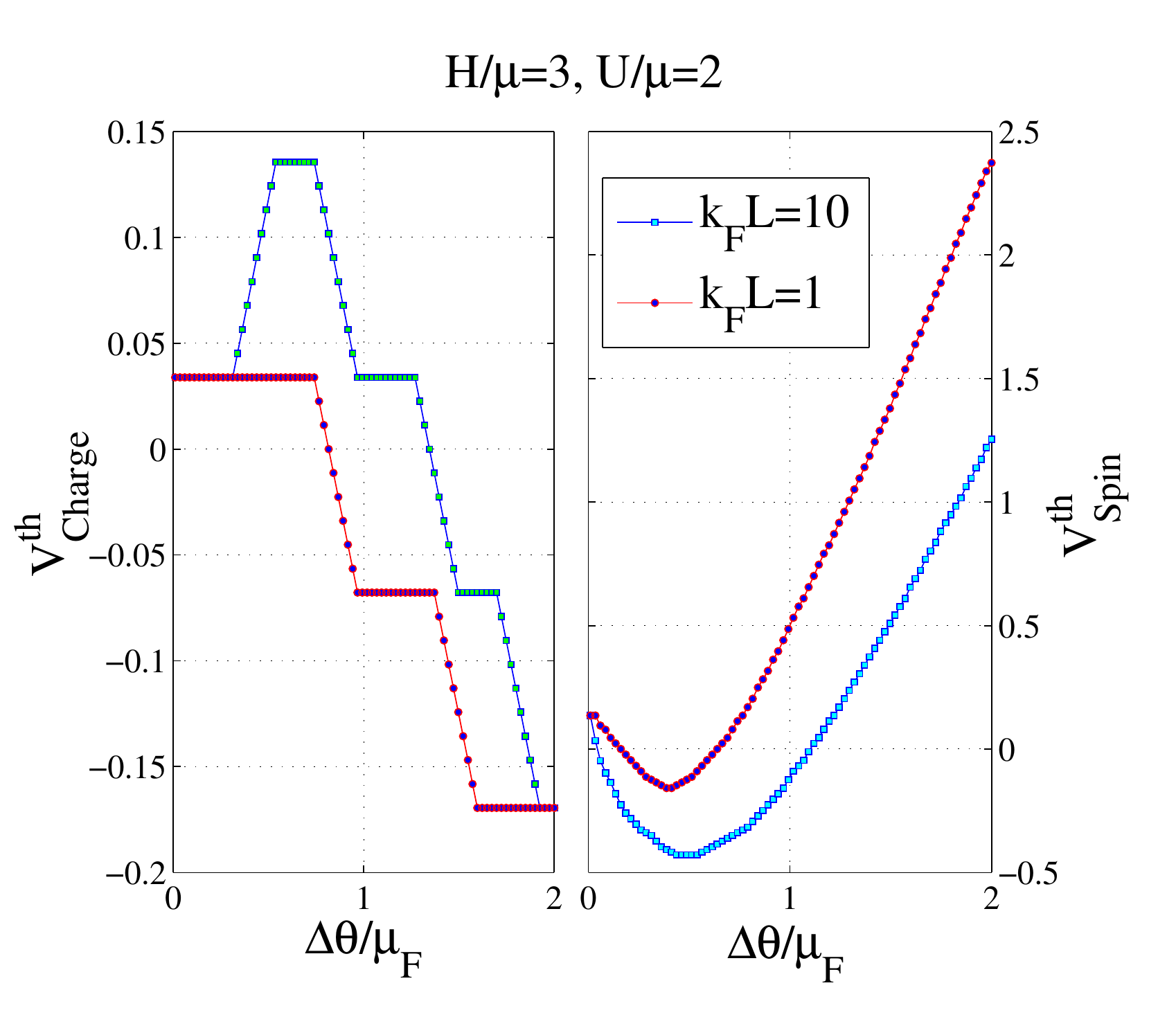}
 \caption{(Color online) Charge (left panel) and spin (right panel) thermovoltage  as functions of temperature difference $\Delta\theta$. We fixed the dimensionless gate voltage and exchange field as $U/\mu_F=2$ and  $H/\mu_F=3$, respectively. }
\label{fig8}
\end{figure}

Now we turn to the nonlinear regime. The thermovoltage can be determined from open-circuit condition. Then, we define charge and spin thermovoltages which can be obtained as $I_{Charge}(V^{th}_{Charge},\Delta\theta)=0$ and $I_{Spin}(V^{th}_{Spin},\Delta\theta)=0$, respectively. Solving this equation, we find the charge $V^{th}_{Charge}$ and spin $V^{th}_{Spin}$ thermovoltages. Results are presented in Fig.~\ref{fig8}.  However, before going through the results of thermovoltages, it would be instructive to analysis the possible thermally activated current (thermocurrent) in the structure. The charge and spin currents are defined as $$I^{th}_{Charge}(V,\Delta\theta)=(I_{+}+I_{-})/2$$ and $$I^{th}_{Spin}(V,\Delta\theta)=(I_{+}-I_{-})$$ where $V$ is biased voltage applied across the junction which we put zero $(V=0)$ in our calculations to take consider just thermally excited flow of charge carriers. We depict our numerical results for some parameters in Fig.~\ref{fig7}.  With increasing   $\Delta\theta/\mu_F$, for case with $k_FL=1$ (see right column) both charge and spin currents magnitude increase in a nonlinear fashion. In case with $K_FL=10$ (see left column) situation is different.  Depends on the exchange field magnitude, shifts the up and down spin subbands on conduction and valance bands or leave them on the conduction band, charge and spin currents show a minimum value at intermediate $\Delta\theta/\mu_F$. Further increase of the temperature difference reduces the current magnitude and reach zero. Subsequent growth of the temperature difference leads to emergence of current with the reversed polarity.
  \par
Having the current, here thermocurrent,  one can find the thermovoltage with solving $I^{th}(V^{th},\Delta\theta)=0$. Results for case in which the exchange filed is grater than the gate voltage $H>U$ are plotted in Fig.~\ref{fig8}. As it can be seen, $V^{th}_{Charge}$ and $V^{th}_{Spin}$ are showing an opposite response to the temperature difference in such a way that with increasing $\Delta\theta$, charge(spin) thermovoltage increases and reaches maximum(minimum). Further increase of the $\Delta\theta$  reduces thermovoltages till reach zero magnitude $(V^{th}=0)$ at a certain value of temperature difference $\Delta\theta$.  Subsequent growth of the $\Delta\theta$ leads to emergence of the thermovoltages with the reversed polarity.  
  \par
The thermovoltage reversal polarity would  be explained as follows. Let's assume that chemical potentials of the electrodes in the unbiased system $\mu_L=\mu_R=0$, and the charge carriers are electrons, temperature difference in across the junction leads to the flow of electrons from the left (hot) to the right (cool) electrode. To suppress this thermally induced current, a negative thermovoltage establishes which grows in magnitude as $\Delta\theta$ increases. Meanwhile, as the $\Delta\theta$ rises, the Fermi distribution function shape of the left (hot) electrode is getting partially smoothed out. This opens the way for holes to flow to the right(cool) electrode. At a certain value of temperature differential the hole flux completely compensates the electron flux. So, at this value of temperature difference, the thermally induced electric current disappears at $V^{th}=0$. As it can be seen, for shorter junction the point which thermovoltage find zero magnitude, shifts to higher temperature difference. 
\par
Before concluding, we comment on the realization of our funding. Setting $\mu_F=1 meV$ and considering $k_FL=1$ and $k_FL=10$, needs a ferromagnetic junction with lengths around $1\mu m$ and $0.1\mu m$, respectively. In which both lengths are smaller than the spin relaxation length and feasible in present experimental devices\cite{a34}. The exchange field and chemical potential are tunable by in-plane external magnetic field\cite{a35,a36} and external gate \cite{a37}, respectively.  Based on our choose $\mu_F=1 meV$,  the values of $U$ and $H$ in the range of $1-10 meV$ is required which is accessible by the current experimental apparatus.
\section{CONCLUSION}\label{sec4 }
In summary, using wave function matching approach and employing the Landauer-Buttiker formula a ferromagnetic graphene junction with temperature gradient across the system, is studied.  We calculate the thermally excited charge and spin current as well as the thermoelectric voltage (Seebeck effect).  We have found the system under consideration is sensitive to temperature and system length. Different lengths  considered (namely $L=0. 1\mu m$ and  $L=1\mu m$) here, are studied in ballistic (quantum) regime and a profound effects are obtained. Albeit, it seems for length $L=1\mu m$, the diffusion processes should be taken into account. So much work is required to study the system behavior in the crossing from the diffusive regime to the quantum regime. But, as long as the phonon contribution is not important in the Seebeck effect, we do not expect a profound effect will emerge.
 \par
Our calculation also revealed that owing to the electron-hole symmetry the charge Seebeck coefficient is, for an undoped magnetic graphene,  an odd function of chemical potential while the spin Seebeck coefficient is an even function regardless of the temperature gradient and junction length. Another important characteristic of thermoelectric transport, thermally excited current in the nonlinear regime, is examined.  It would be our main finding that with increasing thermal gradient applied to the junction the spin and charge thermovoltages decrease and even become zero for non zeros temperature bias.  We have also found with an accurate tuning external parameter, namely the exchange filed and gate voltage, the temperature gradient across the junction drives a pure spin current without accompanying the charge current.

\end{document}